\documentclass[12pt]{article}
\usepackage{graphicx}
\def \b{{\cal B}}
\def \bea{\begin{eqnarray}}
\def \beq{\begin{equation}}
\def \eea{\end{eqnarray}}
\def \eeq{\end{equation}}
\def \ite{{\it et al.}}
\def \s{\sqrt{2}}

\def \sx{\sqrt{6}}
\def \vc{\vec{C}}
\def \vp{\vec{P}}
\def \vt{\vec{T}}
\begin{document}
\renewcommand{\thetable}{\Roman{table}}
\baselineskip 18pt
\begin{flushright}
TECHNION-PH-2005-4\\
EFI-05-01\\
hep-ph/0503131 \\
March 2005 \\
\end{flushright}

\vskip 2cm
\centerline{\bf THE $b \to s$ PENGUIN AMPLITUDE IN CHARMLESS $B \to PP$
DECAYS}
\bigskip
\centerline{Michael Gronau$^1$ and Jonathan L. Rosner$^2$}
\medskip
\centerline{\it $^1$Department of Physics, Technion--Israel Institute of
Technology,}
\centerline{\it Technion City, 32000 Haifa, Israel}
\medskip
\centerline{\it $^2$Enrico Fermi Institute and Department of Physics,
university of Chicago}
\centerline{\it 5640 South Ellis Avenue, Chicago, IL 60637, USA}
\bigskip
\centerline{\large Abstract}
\bigskip
The $b \to s$ penguin amplitude affects a number of $B$ meson decays to two
pseudoscalar ($P$) mesons in which potential anomalies are being watched
carefully, though none has yet reached a statistically compelling level.
These include (a) a sum of rates for $B^0 \to K^0 \pi^0$ and $B^+ \to K^+
\pi^0$ enhanced relative to  half the sum for $B^0 \to K^+ \pi^-$
and $B^+ \to K^0 \pi^+$, (b) a time-dependent CP asymmetry parameter $S$
for $B^0 \to K^0 \pi^0$ which is low in comparison with the expected value of
$\sin 2 \beta \simeq 0.73$, and (c) a similar deviation in the parameter $S$
for $B^0 \to \eta' K_S$.  These and related phenomena involving vector mesons
in the final state are discussed in a unified way in and beyond the Standard
Model. Future experiments which would conclusively indicate the presence of new
physics are identified.  Several of these involve decays of the strange $B$
meson $B_s$.  In the Standard Model we prove an approximate sum rule for CP
rate differences in $B^0\to K^+\pi^-,~B^+\to K^+\pi^0$ and $B^0\to K^0\pi^0$,
predicting a negative sign for the latter asymmetry.
\bigskip

\leftline{PACS numbers:  13.25.Hw, 11.30.Cp, 12.15.Ji, 14.40.Nd}
\newpage

\centerline{\bf I.  INTRODUCTION}
\bigskip

The decays of $B$ mesons to final states consisting of mesons with $u$, $d$,
and $s$ quarks are rich sources of information on the phases and
magnitudes of elements of the Cabibbo-Kobayashi-Maskawa (CKM) matrix and
on possible physics beyond the Standard Model.  Many of these decays are
dominated by an amplitude in which a $b$ quark undergoes a virtual
transition through an intermediate state of a $W^-$ and a quark with
charge 2/3 ($u$, $c$, or $t$) to a strange quark $s$.  This transition
does not occur in vacuum, being eliminated by renormalization, but can
occur in the presence of the chromoelectric field.  Such a transition is known
as a $b \to s$ {\it penguin amplitude} \cite{Ellis}.

The $b \to s$ penguin amplitude has turned out to be significant in many $B$
meson decays.  It is responsible for branching ratios for $B \to K \pi$ decays
of order 1 to $2 \times 10^{-5}$, depending on whether the pion is neutral or
charged.  It also leads to large branching ratios for $B \to K \eta'$ $\sim 7
\times 10^{-5}$.  Several potential anomalies in these modes have attracted
attention, including a sum of rates for $B^0 \to K^0 \pi^0$ and $B^+ \to
K^+ \pi^0$ slightly enhanced with respect to that expected from the sum for
$B^0 \to K^+ \pi^-$ and $B^+ \to K^0 \pi^+$, a time-dependent asymmetry
parameter $S$ for $B^0 \to K^0 \pi^0$ decay which is low with respect to the
expected value of $\sin 2 \beta \simeq 0.73$, and similar discrepancies in $S$
for the decays $B^0 \to \eta' K^0$ and $B^0 \to \phi K_S$.

In the present paper we discuss a framework for describing
new physics in the $b \to s$ penguin amplitude.  We review
the evidence for possible discrepancies from the standard
picture and indicate ways in which these discrepancies, if they exist,
can be sharpened and correlated with other observations.  We also
attempt to estimate the experimental accuracies which would permit
conclusive identification of new physics.  We concentrate on $B \to
PP$ decays, where $P$ is a light pseudoscalar meson.

Section II gives conventions for meson states,
decay amplitudes, CKM matrix elements, and time-dependent CP asymmetries.
Section III is devoted to a parametrization of $B \to K \pi$ amplitudes,
allowing for new physics contributions of the most general form. 
Section IV reviews the pattern of rates for $B \to K
\pi$ and possible effects of the new parametrization, while Section V is
devoted to rate asymmetries in these processes.  Section VI treats the
possible deviation of $S_{K_S \pi^0}$ from its expected value of $\sin 2
\beta \simeq 0.73$.  A similar discussion for $S_{\eta' K_S}$ occupies Section
VII.  The role of $B_s$ decays in sorting out some of these questions is
mentioned in Section VIII.  Section IX lists some related puzzles in $B \to VP$
and $B \to VV$ decays, where $V$ is a vector meson.  Section X notes some
experimental tests for the pattern of deviations from the standard picture
of $b \to s$ penguin-dominated decays.  Section XI concludes.
\newpage

\centerline{\bf II.  CONVENTIONS: STATES, AMPLITUDES, ASYMMETRIES}
\bigskip

We use conventions for states defined in Refs.\ \cite{Gronau:1994rj} and
\cite{Gronau:1995hn}.  Quark model assignments are as usual (e.g., $B^+
= u \bar b,~B^0 = d \bar b$), with the proviso that states with a $\bar
u$ quark are defined with a minus sign (e.g., $B^- = - b \bar u$) for
convenience in isospin calculations.  For a similar reason, a neutral
pion is $\pi^0 = (d \bar d - u \bar u)\sqrt{2}$. 

The CKM matrix $V$ is unitary, implying (e.g.) $V^*_{ub} V_{ud}
+ V^*_{cb} V_{cd} + V^*_{tb} V_{td} = 0$ and a similar relation with
$d \to s$.  We shall make use of these relations in defining
all amplitudes in terms of two combinations of CKM elements.  The
unitarity of $V$ can be depicted in terms of a triangle with
angles $\alpha = {\rm Arg}(-V^*_{tb} V_{td}/V^*_{ub} V_{ud})$,
$\beta = {\rm Arg}(-V^*_{cb} V_{cd}/V^*_{tb} V_{td})$, and
$\gamma = {\rm Arg}(-V^*_{ub} V_{ud}/V^*_{cb} V_{cd})$.

We shall define a set of reduced matrix elements known as tree,
color-suppressed, and penguin amplitudes,  
restricting our attention to strangeness-changing ($|\Delta S| = 1$) processes.
These processes, which were described by primed amplitudes in 
Ref.\ \cite{Gronau:1994rj}, will be presented here as unprimed.
A tree amplitude, $T$, and a color-suppressed amplitude, $C$, involve a 
CKM factor $V^*_{ub}V_{us}$, while a penguin amplitude, $P$, contains
a factor $V^*_{tb}V_{ts} = -V^*_{cb}V_{cs}-V^*_{ub}V_{us} =
-V^*_{cb}V_{cs}(1 + {\cal O}(2\%))$. Color-allowed and
color-suppressed electroweak penguin amplitudes, $P_{EW}$ and $P^c_{EW}$, 
including a CKM factor $V^*_{tb}V_{ts}$, appear with the tree, color-suppressed,
and penguin amplitudes in the independent combinations~\cite{Gronau:1995hn},
\beq
t \equiv T + P^c_{EW},~~c \equiv C + P_{EW},~~
p \equiv P - \frac{1}{3} P^c_{EW}~~~.
\eeq
We will neglect exchange and annihilation amplitudes, $E$ and $A$, 
which are suppressed relative to the dominant $P$ amplitude by  
$|V^*_{ub}V_{us}/V^*_{cb}V_{cs}|(\Lambda_{\rm QCD}/m_b)\sim {\cal 
O}(10^{-3})$~\cite{Gronau:1994rj,Bauer:2004ck}.
A small isosinglet penguin-annihilation amplitude, $PA$, will be 
absorbed in the definition of $P$. Of these three amplitudes only $A$ 
contributes in $B\to K\pi$, while $E$ and $PA$ occur in $B_s\to K\bar K$ 
and $B_s\to\pi\pi$.

CP-violating decay asymmetries are defined as
\beq
A_{CP}(B \to f) \equiv \frac{\Gamma(\bar B \to \bar f) - \Gamma(B
\to f)}{\Gamma(\bar B \to \bar f) + \Gamma(B \to f)}~~,
\eeq
while CP-averaged decay rates are defined by
\beq
\bar \Gamma(f) \equiv \frac{\Gamma(B \to f) + \Gamma(\bar B \to \bar f)}
{2}~~~.
\eeq
For decay to a CP eigenstate $f$, one can measure
time-dependent asymmetry parameters $A_f$ and $S_f$ which occur in
the expression
\beq
\frac{\Gamma(\bar B^0(t) \to f) - \Gamma(B^0(t) \to f)}
{\Gamma(\bar B^0(t) \to f) + \Gamma(B^0(t) \to f)} =
A_f \cos \Delta m t + S_f \sin \Delta m t~~~.
\eeq
Here $\Delta m \simeq 0.5$ ps$^{-1}$ is the mass difference between
neutral $B$ mass eigenstates, while $B^0(t)$ or $\bar B^0(t)$ denotes
a time-evolved state which has been identified as a $B^0$ or $\bar
B^0$ at proper time $t=0$.  One sometimes sees also the notation
$C_f = - A_f$.  The time-integrated rate asymmetry $A_{CP}(f)$ is
equal to $A_f$.

\bigskip

\centerline{\bf III.  PARAMETRIZATION OF $B \to K \pi$ DECAY  AMPLITUDES}
\bigskip

The four $B \to K \pi$ decay amplitudes may be written in a standard
flavor-SU(3) decomposition \cite{Gronau:1994rj,Gronau:1995hn} as
\bea 
A(B^+ \to K^0 \pi^+) & = & p ~~, \label{eqn:0+}\\
A(B^+ \to K^+ \pi^0) & = & -(p+t+c)/\s ~~, \label{eqn:+0}\\
A(B^0 \to K^+ \pi^-) & = & -(p+t) ~~, \label{eqn:+-}\\
A(B^0 \to K^0 \pi^0) & = & (p-c)/\s \label{eqn:00}~~.
\eea 
They satisfy an isospin sum rule \cite{Lipkin:1991st}
\beq \label{eqn:sumrule}
A(B^+ \to K^0 \pi^+) + \s A(B^+ \to K^+ \pi^0) = A(B^0 \to K^+ \pi^-)
 + \s A(B^0 \to K^0 \pi^0)
\eeq
which is a consequence of there being only three independent amplitudes
(two with $I(K \pi) = 1/2$ and one with $I(K \pi) = 3/2$) to describe
the four processes.  The linear combinations shown are those with $I(K
\pi) = 3/2$.

Motivated by early suggestions that the $b \to s$ penguin amplitude was
a promising source of effects due to new physics
\cite{Gronau:1996rv,Grossman:1996ke,London:1997zk,Grossman:1999av},
many modifications of it have been proposed~\cite{NPmodels}.
We shall consider the case of separate new physics operators for 
$b \to s u \bar u$, $b \to s d \bar d$, and $b
\to s s \bar s$ transitions, denoted by $\Delta P_u$, $\Delta P_d$, and 
$\Delta P_s$, with a superscript $(^c)$ to denote those transitions in which 
members of the light $q \bar q = u \bar u,~d \bar d,~s \bar s$ pair end up in 
different mesons.  These will be seen to resemble electroweak penguin
terms, though they could arise from a variety of new-physics sources. 

The $B \to K \pi$ decay amplitudes then may be written in the form
of Eqs.\ (\ref{eqn:0+})--(\ref{eqn:00}) with the identifications
\bea
p & = & P - \frac{1}{3}P^c_{EW} +\Delta P^c_d ~~. \label{eqn:newp} \\
t & = & T + P^c_{EW} + \Delta P^c_u - \Delta P^c_d ~~, \label{eqn:newt} \\
c & = & C + P_{EW} + \Delta P_u - \Delta P_d~~, \label{eqn:newc} 
\eea
The non-electroweak penguin part of $p$  is identified with $\Delta P^c_d$,
while the amplitudes $t$ and $c$ acquire new pieces $\Delta P^c_u - \Delta P^c_d$ and
$\Delta P_u - \Delta P_d$, respectively.
Whereas the standard model amplitude $P$ behaves like an isosinglet
and is therefore common (up to a factor $\s$) to all four $B\to K\pi$ decays,
this is not necessarily the case for the new physics amplitudes which
generally obey $\Delta P^{(c)}_u \ne \Delta P^{(c)}_d$.  It is convenient to 
classify potential anomalies in $B\to K\pi$ in terms of the new physics 
amplitudes $\Delta P^{(c)}_{u,d}$.  For instance, the term $\Delta P^c_d$ 
would show up as a CP asymmetry in $B^{\pm}\to K\pi^{\pm}$, assuming in 
general that this term involves strong and weak phases which differ from
those of $P- P^c_{EW}/3$. In Sections IV and VI we will give 
examples for signatures characterizing the other three terms.
Note that the isospin quadrangle relation (\ref{eqn:sumrule}) of course continues to hold
as long as one assumes that new physics is given by four-quark 
$b\to sq\bar q$ operators implying the absence of $\Delta I > 1$ transitions.

\bigskip

\centerline{\bf IV.  PATTERN OF $B \to K \pi$ RATES}
\bigskip

Current averages for branching ratios for $B \to K \pi$ decays
\cite{Group:2004cx} are quoted in Table \ref{tab:Kpiavs}.  To compare decay
rates one also needs the lifetime ratio $\tau_+/\tau_0 \equiv
\tau(B^+)/\tau(B^0)$, for which the
latest average \cite{Group:2004cx} is $1.081\pm0.015$.  CP-violating
asymmetries are also quoted for use in Sec.\ V.

\begin{table}
\caption{CP-averaged branching ratios (in units of $10^{-6}$) and CP
asymmetries $A_{CP}$ (see Sec.\ V) for $B \to K \pi$ decays.
\label{tab:Kpiavs}}
\begin{center}
\begin{tabular}{c c c} \hline \hline
Decay mode & Branching ratio & $A_{CP}$ \\ \hline
$B^+\to K^0 \pi^+$ & $24.1\pm1.3$ & $-0.020\pm0.034$ \\
$B^+\to K^+ \pi^0$ & $12.1\pm0.8$ &  $0.04 \pm 0.04$ \\
$B^0\to K^+ \pi^-$ & $18.2\pm0.8$ & $-0.109\pm0.019$ \\
$B^0\to K^0 \pi^0$ & $11.5\pm1.0$ & $-0.08\pm0.14$ \\ \hline \hline
\end{tabular}
\end{center}
\end{table}

In the Standard Model the four $B\to K\pi$ amplitudes are dominated by the 
amplitude $p$. Expanding decay rates in $|t/p|$ and $|c/p|$, one observes a
simple sum rule for $B$ decay rates, which holds 
to first order in these ratios~\cite{Gronau:1998ep,Lipkin:1998ie},
\beq
2\Gamma(B^+\to K^+\pi^0) + 2\Gamma(B^0\to K^0\pi^0) = 
\Gamma(B^+\to K^0\pi^+) + \Gamma(B^0\to K^+\pi^-)~~.
\eeq
In terms of specific contributions, this reads
\beq
2|p|^2 + 2{\rm Re}(p^*t) +|t|^2 + 2|c|^2 + 2{\rm Re}(c^*t) = 2|p|^2 + 2{\rm Re}
(p^*t) + |t|^2~~.
\eeq
A similar sum rule holds for $\bar B$ decay rates and for CP-averaged rates. 
Thus,
\beq\label{SR}
2\bar\b(B^+\to K^+\pi^0) + 2\frac{\tau_+}{\tau_0}\bar\b(B^0\to K^0\pi^0)
=\bar\b(B^+\to K^0\pi^+) + \frac{\tau_+}{\tau_0}\bar\b(B^0\to K^+\pi^-)~~,
\eeq
where $\bar \b$ denotes a CP-averaged branching ratio.
Using experimental values for branching ratios and for the lifetime ratio, 
this sum rule reads in units of $10^{-6}$
\beq
(24.2 \pm 1.6) + (24.9 \pm 2.2) = (24.1 \pm 1.3) + (19.7 \pm 0.9)~~,
\eeq
or
\beq
49.1 \pm 2.7 = 43.8 \pm 1.6~~.
\eeq
The two sides differ by $5.3 \pm 3.1$, or $(12 \pm 7)\%$ 
of the better known right-hand-side. This fraction is given to leading order by 
second order terms, ${\rm Re}\langle{c^*(c +t)}\rangle/|p|^2$, where an average 
is taken over $B$ and $\bar B$ contributions. Typical estimates of these terms 
(see, e.g., \cite{Gronau:2003kj}) in the Standard Model limit them to no more 
than a few percent. Fits based on flavor SU(3) \cite{Chiang:2004nm,Suprun:2005}
predict branching ratios satisfying Eq.\ (\ref{SR}) more accurately, 
obtaining a slightly smaller value of $\bar\b(B^0\to K^0\pi^0)$ than observed
due to destructive interference between the two dominant terms contributing to 
this process, $(P-\frac{1}{3}P^c_{EW})/\s$ and $-P_{EW}/\s$.

An equivalent approach to the sum rule can be presented in terms of an equality 
between two ratios of CP-averaged branching ratios (equivalently, of decay
rates) defined as~\cite{Buras:1998rb}
\beq
R_c \equiv \frac{2 \bar\b(B^+ \to K^+ \pi^0)}{\bar\b(B^+ \to K^0 \pi^+)}~~,~~~
R_n \equiv \frac{\bar\b(B^0 \to K^+ \pi^-)}{2 \bar\b(B^0 \to K^0 \pi^0)}~~.
\eeq
The experimental values are
\beq
R_c = 1.00 \pm 0.09~~,~~~R_n = 0.79 \pm 0.08~~,~~~R_c - R_n = 0.21
\pm 0.12~~.
\eeq
Expanding $R_c$ and $R_n$ in ratios $t/p, c/p$ and their charge conjugates, 
one can show that the difference $R_c - R_n$ is quadratic in these ratios.
Attention has been called \cite{Buras:2003yc,Buras:2003dj,Buras:2004ub,
Buras:2004th} to the fact that  if the difference $R_c - R_n$ is maintained 
with improved statistics this could signal new physics.

In the absence of differences between penguin terms $\Delta P_u$ and $\Delta P_d$
or $\Delta P^c_u$ and $\Delta P_c^d$, one would most naturally ascribe a large
term of the form ${\rm Re}\langle{c^*(c +t)}\rangle/|p|^2$ to color-favored
electroweak penguin terms \cite{Yoshikawa:2003hb} of magnitude
larger than expected.  The point we wish to stress here is that
any four-quark operator which contributes to $\Delta P_u - \Delta P_d$ will emulate
the color-allowed electroweak penguin $P_{EW}$ in Eq.\ (\ref{eqn:newc}), 
while any four-quark operator which contributes to $\Delta P^c_u - \Delta P^c_d$
will emulate $P^c_{EW}$ in Eq.\ (\ref{eqn:newt}).  Thus, both $c$
and $t$ can receive contributions from new physics aside from
enhanced electroweak penguins as long as $b \to s q \bar q$ operators
produce $u \bar u$ pairs differently from $d \bar d$ pairs.

It is interesting to note, as has been pointed out \cite{Ligeti:2004ak},
that the Fleischer-Mannel ratio~\cite{Fleischer:1997um},
\beq
R \equiv \frac{\bar\Gamma(B^0 \to K^+ \pi^-)}{\bar\Gamma(B^+ \to K^0 \pi^+)}~~,
\eeq
is currently 
\beq
R = 0.816 \pm 0.058~~,
\eeq
differing from 1 by $3.2 \sigma$.  At 95\% confidence level $R < 0.911$. 
Neglecting $P^c_{EW}$  terms, this would lead through the Fleischer-Mannel
bound $\sin^2 \gamma \le R$ to an upper limit $\gamma \le
73^\circ$. 

However, as we mention in the next Section, $P^c_{EW}$
and $C$ are not much suppressed relative to $P_{EW}$ and $T$, 
respectively, as has been customarily assumed. 
Including the $P^c_{EW}$ term,  the Fleischer-Mannel bound 
becomes~\cite{Gronau:1997an}
\beq
\sin^2\gamma \le \frac{R}{|1 + P^c_{EW}/(P-\frac{1}{3}P^c_{EW})|^2} \approx
|1- \frac{P^c_{EW}}{P}|^2R~~.
\eeq
The effect on the bound depends on the magnitude of $P^c_{EW}/P$, which is 
typically a few percent, and on the phase of this ratio. Using, for instance,
Fit III in~\cite{Chiang:2004nm} based on SU(3), one finds central values 
$|P^c_{EW}/P| = 0.044, {\rm Arg}(P^c_{EW}/P)=-69^\circ$, implying 
$\sin^2\gamma \le 0.97R \le 0.884$, or $\gamma \le 70^\circ$. This upper
bound is consistent with other Standard Model constraints on 
$\gamma$~\cite{CKMfitter,UTfit}.
A potential inconsistency would have been ascribed to $\Delta P^c_u$
or $\Delta P^c_d$, which occur in $p+t$ and $p$, respectively. 

\bigskip

\centerline{\bf V.  RATE ASYMMETRIES IN $B \to K \pi$}
\bigskip

The penguin dominance of the $B \to K \pi$ decay amplitudes was used
in Ref.\ \cite{Gronau:1998ep} to derive in the Standard Model a relation 
between direct CP-violating rate differences in various $B \to K \pi$ processes.
The simplest of these was based on assuming that the only important
amplitude interfering with $p$ was $t$, in which case the relation
\beq \label{eqn:simplest}
\Delta (K^+ \pi^-) \sim 2 \Delta (K^+ \pi^0)
\eeq
was obtained.  Here
\bea
\Delta (K^+ \pi^-) & \equiv & \Gamma(B^0 \to K^+ \pi^-) - \Gamma(\bar
B^0 \to K^- \pi^+)~~, \\
\Delta (K^+ \pi^0) & \equiv & \Gamma(B^+ \to K^+ \pi^0) - \Gamma(B^-
\to K^- \pi^0)~~,
\eea
with similar definitions for $\Delta(K^0 \pi^+)$ and $\Delta(K^0
\pi^0)$.  These rate asymmetries are related to the CP asymmetries
as defined in Sec.\ II
by $\Delta(f) = - 2 A_{CP}(f) \bar \Gamma(f)$, where the CP-averaged rate
$\bar \Gamma(f)$ was defined in Sec.\ II.

Since $\bar \Gamma(K^+ \pi^-) \approx 2 \bar \Gamma(K^+ \pi^0)$, the
relation (\ref{eqn:simplest}) reduces to the prediction
\beq \label{eqn:simpleACP}
A_{CP}(B^0 \to K^+ \pi^-) \sim A_{CP}(B^+ \to K^+ \pi^0)~~,
\eeq
which is rather far from what is observed.  According to the averages in
Table \ref{tab:Kpiavs}, the left-hand side of Eq.\ (\ref{eqn:simpleACP})
is $-0.11 \pm 0.02$, while the right-hand side is $0.04 \pm 0.04$.
The two sides thus differ by more than $3 \sigma$.  Is this a problem?
Does this indicate isospin-violating new physics~\cite{Hou:2005hd}?

As in Ref.\ \cite{Gronau:1998ep}, we define $2 \vp \vt$ to
be the interference between $P$ and $T$ contributing to the rate
difference $\Delta(K^+ \pi^-)$, with similar notations for other
interference terms and rate differences.  One then finds
\cite{Gronau:1998ep} that
\bea
\Delta(K^0 \pi^+)  & \simeq & 0~~, \\
\Delta(K^+ \pi^0) & \simeq & \vp \vt + \vp \vc + (\vp_{EW} + \frac{2}{3}
\vp^c_{EW}) (\vt+\vc) ~~, \\
\Delta(K^+ \pi^-) & = & 2 \vp \vt + \frac{4}{3} \vp^c_{EW} \vt~~,\\
\Delta(K^0 \pi^0) & = & - \vp \vc + \vp_{EW} \vc +
\frac{1}{3}\vp^c_{EW} \vc~~.
\eea
The only interference terms
which contribute to direct CP-violating rate differences are those
which have differing weak and strong phases.  Thus one sees no
interference between $C$ and $T$ or between electroweak penguin
terms and $P$. 

The relation (\ref{eqn:simplest}) was derived by neglecting all terms
in the rate differences except $\vp \vt$.  An argument was given for
the relative smallness of the term $\vp \vc$ under the assumption
that $|C/T| = {\cal O}(1/5)$.  Recent fits based on flavor SU(3)
\cite{Chiang:2004nm,otherSU3fits,Baek:2004rp} indicate that 
$|C/T|$ is more like 0.7 to 0.9 (quoting the results of fits in 
\cite{Chiang:2004nm} which include processes involving
$\eta$ and $\eta'$ as well as kaons and pions; $|C/T|$ is even 
larger in a fit studying only $B\to K\pi$~\cite{Baek:2004rp}).  
Also, arguments based on 
a Soft Collinear Effective Theory~\cite{Bauer:2004ck,Bauer:2004tj} 
show that $C$ and $T$ are comparable. In this case an
improved relation based on similar reasoning retains the $\vp \vc$
term and is
\beq \label{eqn:better}
\Delta(K^+ \pi^-) \approx 2 \Delta(K^+ \pi^0) + 2 \Delta(K^0 \pi^0)~~.
\eeq
This relation ignores terms on the right-hand side which can be written as
\beq\label{rest}
 (\vp_{EW} + \vp^c_{EW})(\vt + \vc) + (\vp_{EW} \vc - \vp^c_{EW} \vt)
\approx 0~~.
\eeq
An argument for the smallness of the first term was given 
in~\cite{Gronau:1998ep} using a property of the $I(K\pi)=3/2$ amplitude,  
$(T+C)+(P_{EW}+P^c_{EW})$, in which the two terms involve approximately 
a common strong phase~\cite{Neubert:1998pt}.
The second term in (\ref{rest}) vanishes approximately due to a relation 
$P^c_{EW}/P_{EW} \approx C/T$~\cite{Gronau:1998fn}. Both approximations 
follow from flavor SU(3) when neglecting electroweak penguin operators with
small Wilson coefficients ($c_7$ and $c_8$).

Using the approximate relations $\bar \Gamma(K^+ \pi^-) \simeq  2 \bar
\Gamma(K^+ \pi^0) \simeq 2 \bar \Gamma(K^0 \pi^0)$, Eq.\
(\ref{eqn:better}) may be transcribed as
\beq\label{A_Kpi}
A_{CP}(K^+ \pi^-) \approx A_{CP}(K^+ \pi^0) + A_{CP}(K^0 \pi^0)~~,
\eeq
which reads, according to Table \ref{tab:Kpiavs}, as
\beq
-0.109 \pm 0.019 \approx (0.04 \pm 0.04) + (-0.08 \pm 0.14)~~~.
\eeq
Another way to put this relation is that $A_{CP}(K^0 \pi^0)$ is
predicted to be $-0.15 \pm 0.04$, i.e., non-zero at a level greater
than $3 \sigma$.  A more precise prediction, using the measured rates of the
above three processes, is $A_{CP}(K^0\pi^0) = - 0.13 \pm 0.04$.
CPT requires that the overall direct CP asymmetry 
vanishes in eigenstates of the strong S matrix. Our prediction excludes
the possibility that the asymmetry in $B^0\to K^0\pi^0$ alone 
compensates for the observed asymmetry in 
$B^0\to K^+\pi^-$~\cite{Lipkin:1995hn}. The two asymmetries are 
predicted to have equal signs.

In Ref.~\cite{Gronau:2003kx} we noted a relation between CP rate-differences, 
which holds in the limit of SU(3) when neglecting annihilation-like amplitudes
($E + PA$) in $B^0\to\pi^0\pi^0$,
\beq
\Delta(\pi^0\pi^0) = - \Delta(K^0\pi^0)~~,
\eeq
or
\beq
A_{CP}(\pi^0\pi^0) = 
-\frac{\bar\b(B^0\to K^0\pi^0)}{\bar\b(B^0\to\pi^0\pi^0)} A_{CP}(K^0\pi^0)~~.
\eeq
Using our prediction, $A_{CP}(K^0\pi^0)=-0.13 \pm 0.04$, and the two branching
ratios~\cite{Group:2004cx}, 
$\bar\b(B^0\to K^0\pi^0) = (11.5 \pm 1.0)\times 10^{-6},~
\bar\b(B^0\to \pi^0\pi^0) = (1.45 \pm 0.29)\times 10^{-6}$, 
we find
\beq
A_{CP}(\pi^0\pi^0) = +1.0 \pm 0.4~~.
\eeq
This {\it large and positive} value should be compared with the current
world-averaged value~\cite{Group:2004cx}, $A_{CP}(\pi^0\pi^0) = +0.28\pm0.39$.
It would 
be interesting to watch the decrease of experimental errors in order to learn 
the effects of SU(3) breaking corrections and annihilation-like amplitudes.

\bigskip

\centerline{\bf VI.  DEVIATIONS OF $S_{K \pi}$ FROM ITS NOMINAL VALUE}
\bigskip

The dominance of the $b \to s$ penguin amplitude in $B^0 \to K^0 \pi^0$
implies that the parameter $S_{K \pi} \equiv S_{K_S \pi^0}$ should be very
close to the value $\sin(2 \beta)$ expected from interference between 
$B^0$--$\bar B^0$ mixing and $B^0$ decay alone.  
One has 
\beq
S_{K \pi} = \frac{2 {\rm Im} \lambda_{K \pi}}{|\lambda_{K \pi}|^2+1}
~~,~~~A_{K \pi} = \frac{|\lambda_{K \pi}|^2-1}{|\lambda_{K \pi}|^2+1}~~, 
\eeq
where
\beq
\lambda_{K \pi} \equiv - e^{- 2 i \beta} \frac{A(\bar B^0 \to \bar
K^0 \pi^0)}{A(B^0 \to K^0 \pi^0)}~~~.
\eeq
Rewriting Eq.\ (\ref{eqn:00}) for $A(B^0 \to K^0 \pi^0)$  in terms of two 
contributions $A_P$ and $A_C$ involving CKM factors $V^*_{cb}V_{cs}$ 
and $V^*_{ub}V_{us}$, respectively, and a relative strong phase $\delta$,
\beq
A(B^0\to K^0\pi^0) = A_P + A_C = |A_P|e^{i\delta} + |A_C|e^{i\gamma}~~,
\eeq
where
\beq
A_P \equiv  (P-P_{EW} - \frac{1}{3}P^c_{EW})/\s~,~~~~A_C \equiv -C/\s~~,
\eeq
one obtains to first order in $|A_C/A_P|$~\cite{Gronau:1989ia}
\beq
\Delta S_{K \pi} \equiv S_{K \pi} - \sin 2 \beta \approx 2|A_C/A_P|
\cos 2 \beta \cos \delta \sin \gamma~~,~~~
A_{K \pi} \simeq -2 |A_C/A_P| \sin \delta \sin \gamma~~~.
\eeq

\begin{table}
\caption{Time-dependent CP asymmetry parameters for $B^0 \to K_S \pi^0$.
\label{tab:Kpiasym}}
\begin{center}
\begin{tabular}{c c c c} \hline \hline
Parameter & BaBar~\cite{BaBarKpi} & Belle~\cite{BelleKpi} & Average \\ \hline
$S_{K \pi}$ & $0.35^{+0.30}_{-0.33}\pm0.04$ & $0.32\pm0.61\pm0.13$ &
 $0.34^{+0.27}_{-0.29}$ \\
$A_{K \pi}$ & $-0.06\pm0.18\pm0.03$ & $-0.11\pm0.20\pm0.09$ &
 $-0.08 \pm 0.14$ \\ \hline \hline
\end{tabular}
\end{center}
\end{table}

With the help of information on the $B^0 \to \pi^0 \pi^0$ decay rate
and an upper limit on $\bar\b(B^0 \to K^+ K^-)$, it was found
(using flavor SU(3)) that~\cite{Gronau:2003kx}
\beq
-0.11 \le \Delta S_{K \pi} \le 0.12~~,~~~|A_{K \pi}| \le 0.17~~,
\eeq
under the assumption that $A(B^0 \to K^+ K^-)$ could be neglected, or
\beq
-0.18 \le \Delta S_{K \pi} \le 0.16~~,~~~|A_{K \pi}| \le 0.26~~,
\eeq
when taking into account a possible non-zero amplitude for $B^0 \to
K^+ K^-$.  
[These constraints are modified slightly by recent updates of
$\bar\b(B^0\to K^0\pi^0)$
and $\bar\b(B^0\to\pi^0\pi^0)$~\cite{Group:2004cx}.]
Under the first, more restrictive, assumption one could
actually {\it exclude} a small elliptical region in the $S_{K \pi},
A_{K \pi}$ plane with center at $(0.76,0)$ and semi-axes
$(0.06,0.08)$. Our prediction (\ref{A_Kpi}) of a negative direct asymmetry
is consistent with these bounds and implies $\sin\delta > 0$.

The current experimental situation for measurements of $S_{K \pi}$ and
$A_{K \pi}$ is summarized in Table \ref{tab:Kpiasym}.  The observed
$\Delta S_{K \pi} = -0.39^{+0.27}_{-0.29}$ differs from zero by $1.4
\sigma$.  If one were to ascribe this difference to non-standard
behavior of the $b \to s q \bar q$ penguin amplitude, one would have
to blame the amplitude $\Delta P^c_d$ or $\Delta P_u-\Delta P_d$, 
modifying respectively the $p$ or $c$ amplitude.  At this point, however, 
it is clearly premature to speculate on such modifications. 

A flavor SU(3) fit to $B \to PP$ amplitudes \cite{Chiang:2004nm}
predicts a {\it positive} $\Delta S_{K \pi} \simeq 0.1 \pm 0.01$ as well as a 
negative $A_{K \pi} \simeq -0.12 \pm 0.03$. The latter prediction is in accord 
with the discussion of the previous section.  The sign of the former may be
understood from the following qualitative argument. In SU(3) fits the two
terms, $p=P-\frac{1}{3}P^c_{EW}$ and $P_{EW}$, are found to involve a relative
strong phase smaller than $\pi$ and thus interfere destructively in $A_P$.  To
account for the somewhat large measured CP-averaged rate of $B^0\to K^0\pi^0$,
which is equal to half the rate of $B^+\to K^0\pi^+$ given by $p$ alone, this
requires constructive interference between $A_P$ and $A_C$ in the CP-averaged
rate for $B^0 \to K^0 \pi^0$. This implies $\cos\delta > 0$ and consequently 
$\Delta S_{K \pi} > 0$.
\bigskip

\centerline{\bf VII.  DEVIATIONS OF $S_{\eta' K_S}$ FROM ITS NOMINAL VALUE}
\bigskip

The experimental situation with regard to the time-dependent parameter
$S_{\eta' K}$ for $B^0 \to \eta' K^0$ is not clear.  The BaBar
Collaboration sees a significant deviation from the standard picture
prediction of $\sin 2 \beta \simeq 0.73$, while Belle's value is
consistent with the standard picture.  The values of $S_{\eta' K}$
and $A_{\eta' K}$ and their averages are summarized in Table
\ref{tab:etapKasym}.  Here, in view of the discrepancy between Belle
and BaBar values, we have multiplied the error (as quoted in Ref.\
\cite{Group:2004cx}) by a scale factor $S = \sqrt{\chi^2}$, where
$\chi^2$ is the value for the best fit to the BaBar and Belle values.

\begin{table}
\caption{Time-dependent CP asymmetry parameters for $B^0 \to \eta' K_s$.
Errors on averages include scale factor $S = \sqrt{\chi^2}$.
\label{tab:etapKasym}}
\begin{center}
\begin{tabular}{c c c c c} \hline \hline
Parameter & BaBar~\cite{BaBaretaK} & Belle~\cite{BelleKpi} & $S$ & Average \\ \hline
$S_{\eta' K}$ & $0.30\pm0.14\pm0.02$ & $0.65\pm0.18\pm0.04$ & 1.51 &
$0.43 \pm 0.17$ \\
$A_{\eta' K}$ & $0.21\pm0.10\pm0.02$ & $-0.19\pm0.11\pm0.05$ & 2.53 &
 $0.04 \pm 0.20$ \\ \hline \hline
\end{tabular}
\end{center}
\end{table}

The average value of $A_{\eta' K}$ is consistent with zero, while
$S_{\eta' K}$ differs from $\sin 2 \beta = 0.726 \pm 0.037$ by
$\Delta S_{\eta' K} = -0.30 \pm 0.17$, or $1.76 \sigma$.  In
contrast to the case of $B^0 \to K^0 \pi^0$, there are a wide range
of possible contributors to new physics in $b \to s q \bar q$
amplitudes.  In the flavor-SU(3) decomposition of Ref.\
\cite{Chiang:2004nm} the amplitude for $B^0 \to \eta' K^0$ is
\beq
A(B^0 \to \eta' K^0) = (3 p + 4 s + c)/\sx~~~,
\eeq
where $s$ denotes a singlet penguin amplitude contributing mainly to
$\eta'$ production.  It is expressed in terms of a genuine
singlet-penguin term $S$ and an electroweak penguin correction $P_{EW}$ as
$s = S - (1/3)P_{EW}$.  New-physics contributions for $b \to s u \bar
u$ or $b \to s s \bar s$ can enter into the $s$ and $c$ amplitudes,
while those for $b \to s d \bar d$ can enter into all three amplitudes.
Thus, it becomes particularly hard to identify the source of new
physics if the only deviation from the standard prediction for $S$
is that seen in $B^0 \to \eta' K^0$.

A question arises as to the accuracy with which the standard picture can
predict $\Delta S_{\eta' K}$ and $A_{\eta' K}$.  We have addressed this in
two ways in previous work.  (1) In Ref.\ \cite{Gronau:2004hp} we used flavor
SU(3) (or only its U-spin subgroup~\cite{Chiang:2003rb,Grossman:2003qp}) to
bound the effects of non-penguin amplitudes which could give rise to non-zero
$\Delta S_{\eta' K}$ and $A_{\eta' K}$.  (2) In Ref.\ \cite{Chiang:2004nm} we
performed a fit to a wide variety of $B \to PP$ processes based on flavor
SU(3), obtaining predictions for these quantities
\beq \label{eqn:etapKpred}
\Delta S_{\eta' K} \approx 0.02 \pm 0.01~~,~~~
A_{\eta' K} \approx 0.06 \pm 0.02 ~~~.
\eeq
Other explicit calculations (see, e.g., \cite{BN,Cheng:2005bg}) also obtain such
very small values in the standard picture.

While it is difficult to estimate the deviations from Eq.\
(\ref{eqn:etapKpred}) that might cause us to question the standard
picture, one can at least give a range of such deviations that
would {\it not} be a cause for immediate concern.  Proceeding in the
same manner as for $B^0 \to K^0 \pi^0$ in the previous section, we
decompose the amplitude for $B^0 \to \eta' K^0$ into two terms
$A_P$ and $A_C$ (dropping primes in comparison with Ref.\
\cite{Gronau:2004hp}) involving intrinsic CKM factors $V^*_{cb} V_{cs}$
and $V^*_{ub} V_{us}$, and strong and weak phases $\delta$ and
$\gamma$, respectively:
\beq
A(B^0 \to \eta' K^0) = A_P + A_C = |A_P|e^{i \delta} + |A_C|e^{i \gamma}~.
\eeq
First order expressions for $\Delta S_{\eta' K}$ and $A_{\eta' K}$ are the 
same as in $B^0\to K^0\pi^0$:
\beq \label{eqn:etapKdevs}
\Delta S_{\eta' K} \equiv S_{\eta' K} - \sin 2 \beta \approx
2|A_C/A_P| \cos 2 \beta \cos \delta \sin \gamma~~,~~~
A_{\eta' K} \approx - 2|A_C/A_P|\sin \delta \sin \gamma~~~.
\eeq

\begin{figure}
\includegraphics[height=4.9in]{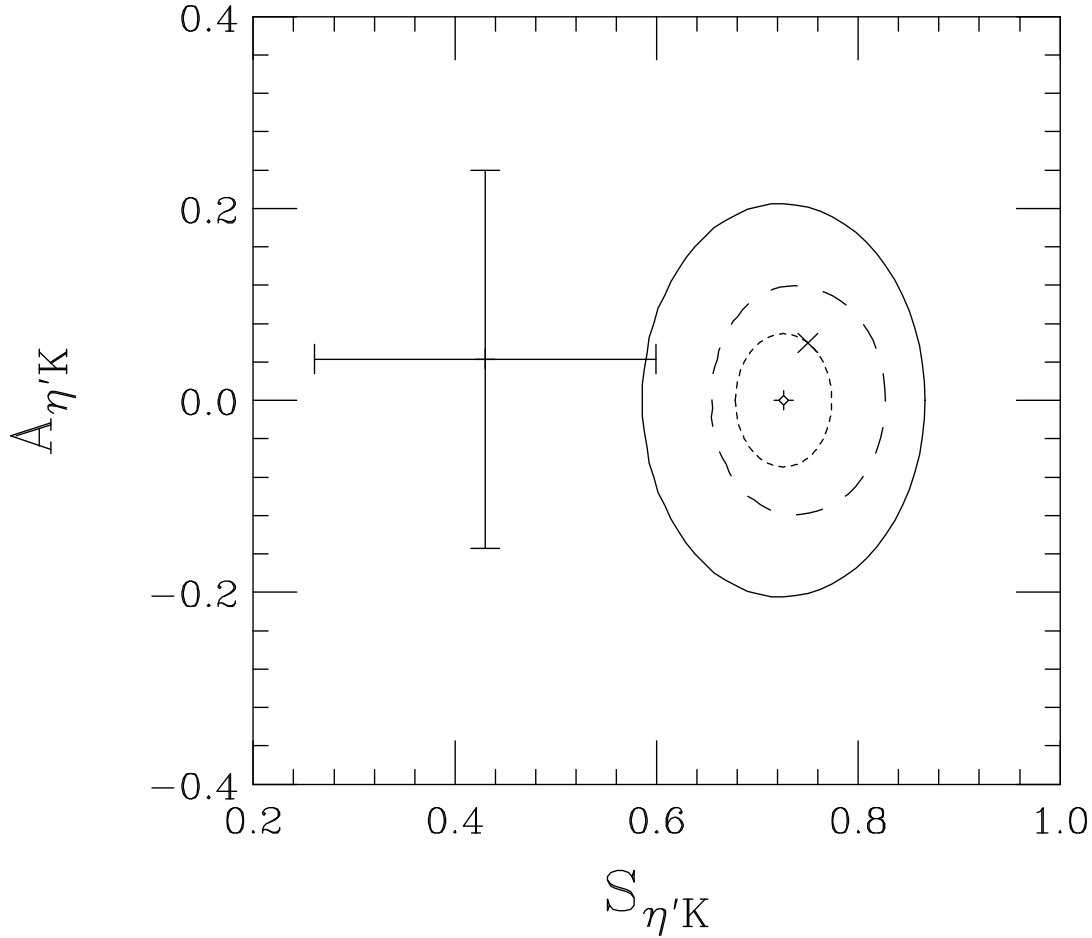}
\caption{Regions in the $S_{\eta' K}, A_{\eta' K}$ plane satisfying
updated limits based on the last line in Table \ref{tab:ant}.  Solid
curve: limits based on flavor SU(3) without neglect of
spectator amplitudes; dashed curve: limits with spectator
amplitudes neglected.  Plotted open point:  $(S_{\eta' K},
A_{\eta' K}) = (0.726,0)$.  Point labeled $\times$:
central value of a prediction in Ref.\ \cite{Chiang:2004nm}.
The dotted ellipse passing through this point denotes the range of values of
$S_{\eta' K},A_{\eta' K}$ in which only the strong phase $\delta$ varies.
\label{fig:etapK}}
\end{figure}

The predictions of Ref.\ \cite{Chiang:2004nm} that $\Delta S_{\eta' K}
\ge 0$, $A_{\eta' K} \ge 0$ do not seem to have a simple interpretation,
in contrast to that for the sign of $\Delta S_{K \pi}$ in the previous section,
since strong phases of several small amplitudes
are involved.  Nonetheless, all the terms in Eq.\ (\ref{eqn:etapKdevs})
with the exception of $\delta$ may be considered to be fairly stable in
the SU(3) fit, so that a crude estimate of possible deviations would
be to let $\delta$ range through all possible values, thereby tracing an
ellipse (shown as the dotted curve in Fig.\ \ref{fig:etapK})
passing through the point (\ref{eqn:etapKpred}).  Indeed, we
would regard any measurement lying {\it within} this ellipse as
providing little challenge to the standard picture, given the
rudimentary nature of our understanding of strong phases.

\begin{table}
\caption{Comparison of current and anticipated 90\% c.l.\ upper limits on
branching ratios (in units of $10^{-6}$) for $B^0$ decays to final
states conisisting of pairs of neutral particles which may be used
to place correlated bounds on $\Delta S_{\eta' K}$ and $A_{\eta' K}$.
\label{tab:ant}}
\begin{center}
\begin{tabular}{c c c c c c c} \hline \hline
Mode    & $\pi^0 \pi^0$ & $\pi^0 \eta$ & $\pi^0 \eta'$ & $\eta \eta$ &
 $\eta' \eta'$ & $\eta \eta'$ \\ \hline
Current & $1.45\pm 0.29$ & 2.5 & 3.7 & 2.0 & 10 & 4.6 \\
Anticipated & 1.8 & 1.5 & 1.6 & 2.0 & 1.2 & 3.3 \\ \hline \hline
\end{tabular}
\end{center}
\end{table}

A more conservative estimate of eventual limits of the standard picture for
$\Delta S_{\eta' K}$ and $A_{\eta' K}$ may be obtained by improving
the bounds set in Ref.\ \cite{Gronau:2004hp} using {\it anticipated}
rather than current upper bounds on strangeness-conserving $B^0$ 
decays to various final states consisting of neutral particles.  In Table
\ref{tab:ant} we compare current bounds~\cite{Group:2004cx} (mostly used in 
Ref.\ \cite{Gronau:2004hp}) with those
that could be set if the data respected 90\% c.l.\ upper limits of
the predictions in the flavor SU(3) fits of Ref.\ \cite{Chiang:2004nm}.

Using the last line in Table \ref{tab:ant} and the current central value
\cite{Group:2004cx,BaBaretaK} for the branching ratio $\b(B^0 \to \eta' K^0)
= 68.6 \times 10^{-6}$, we find a modest improvement in the bounds of Ref.\
\cite{Gronau:2004hp}.  The resulting constraints are illustrated in Fig.\
\ref{fig:etapK}.  The dashed curve denotes SU(3) bounds in which
annihilation-like amplitudes were neglected \cite{Gronau:1994rj,Bauer:2004ck}
as in the discussion of $B\to K\pi$. In that case the previously-excluded
ellipse was centered at (0.74,0) with semi-axes (0.12,0.18).  With the new
inputs the semi-axes shrink to (0.09,0.12). Values of $S_{\eta' K}$
less than 0.65 would cause us first of all to question the neglect of
annihilation-like amplitudes involving the spectator quark.  The absence of a
detectable rate for $B^0 \to K^+ K^-$ \cite{Group:2004cx}, indicating a low
level of rescattering from other states~\cite{Gronau:1998gr}, and a theoretical
argument presented in~\cite{Bauer:2004ck} are the best justifications for their
omission.

To be very conservative, we also present bounds without  neglecting 
annihilation-like amplitudes.  For this case, we found in~\cite{Gronau:2004hp}
values of $S_{\eta' K},A_{\eta' K}$ confined roughly to an ellipse with
center at $(0.71,0)$ and semi-axes $(0.22,0.33)$.  With the
new inputs we now find this ellipse (solid curve in Fig.\ \ref{fig:etapK})
to be centered at $(0.73,0)$ with semi-axes $(0.14,0.20)$.  The lower bound 
on $S_{\eta' K}$ thus becomes 0.59. If the central value of the present average
remains at 0.43 and the error is reduced to $\pm 0.05$, the standard
picture will be in trouble.  This situation is probably some
distance in the future.
\bigskip

\centerline{\bf VIII. THE ROLE OF $B_s$ DECAYS}
\bigskip

A number of decays of strange $B$'s ($B_s \equiv \bar b s$) can be related
to those of non-strange $B$'s using flavor SU(3). In particular, its U-spin
subgroup involving the interchange $d \leftrightarrow s$ relates CP rate
differences in strangeness conserving and strangeness changing decays of $B^0$
and $B_s$ mesons \cite{Fleischer:1999pa,Gronau:2000md,Gronau:2000zy,%
Lipkin:2005}.  A few examples are
\bea
\Delta(B_s\to K^+K^-) & = & -\Delta(B^0\to\pi^+\pi^-)~~,\\
\Delta(B_s\to K^-\pi^+) & = & - \Delta(B^0\to \pi^- K^+)~~,\\
\Delta(B_s\to \bar K^0\pi^0) & = & - \Delta(B^0\to K^0\pi^0)~~.
\eea
Gross violation of these relations, beyond corrections anticipated from SU(3)
breaking, would indicate new physics in $b\to s$ transitions.  As pointed out
in~\cite{Gronau:1996rv}, new physics contributions in these transitions are
often accompanied by anomalous contributions to $B_s-\bar B_s$ mixing, thereby
affecting the $B_s$ mass-difference and time-dependent decays of  $B_s$ meson.
A well-known example is the time-dependent CP asymmetry in $B_s \to K^+K^-$
which is related in the Standard Model to the phase $\gamma$
\cite{Fleischer:1999pa,Gronau:2002bh}.  These measurements would therefore
provide complementary information about potential new-physics operators.

Considering only strangeness changing decays, any anomalous behavior in 
$b \to s$ penguin amplitudes should show up in $B_s$ decays as well as in 
non-strange $B$ decays.  Two of the simplest SU(3) relations, neglecting phase
space and form factor differences and small amplitudes involving the spectator
quark, are~\cite{Gronau:2000zy}
\bea
\Gamma(B_s \to K^+ K^-) & = & |p+t|^2 = \Gamma(B^0 \to K^+ \pi^-)~~, \\
\Gamma(B_s \to K^0 \bar K^0) & = & |p|^2 = \Gamma(B^+ \to K^0 \pi^+)~~.
\eea
These predictions are also obtained in the flavor-SU(3) description of a
wide variety of $B^+$, $B^0$, and $B_s$ decays in Ref.\ \cite{Suprun:2005}.
The first relation becomes a prediction for an approximate equality of
branching ratios under the assumption of equal lifetimes for $B_s$ and $B^0$,
which is consistent with present data \cite{Group:2004cx}.  It is not
particularly well-obeyed since \cite{Warburton:2004vd}
\beq \label{eqn:bkk}
\b(B_s \to K^+ K^-) = (34.3 \pm 5.5 \pm 5.1) \times 10^{-6}~~,
\eeq
to be compared with the world average in Table \ref{tab:Kpiavs}:
\beq
\b(B^0 \to K^+ \pi^-) = (18.2 \pm 0.8) \times 10^{-6}~~.
\eeq
If penguin amplitudes factorize, one may parametrize SU(3) breaking in terms of
a ratio $F^{B_sK}/F^{B\pi}$ of form factors.  The $B_s$ decay rate could be
enhanced to the value (\ref{eqn:bkk}) if this ratio were about 1.4.  Such a
large value was obtained in a calculation based on QCD sum rules
\cite{Khodjamirian:2003xk}. The result (\ref{eqn:bkk}) is still preliminary,
and is based on a fit involving several contributions.  It will be
interesting to see if the value (\ref{eqn:bkk}) persists with improved
statistics and better particle identification capabilities.  No results have
been presented yet for $B_s \to K^0 \bar K^0$, which is difficult to detect in
a hadronic production environment.

The flavor-SU(3) fit of Ref.\ \cite{Suprun:2005} predicts branching ratios
below $10^{-7}$, for $B_s \to \pi^0 (\eta,\eta')$, and large branching ratios
of around $56 \times 10^{-6}$ and $23 \times 10^{-6}$ for $B_s \to \eta' \eta'$
and $B_s \to \eta\eta'$, respectively, partly as a result of a large singlet
penguin contribution.  The rates for $B_s \to \pi^0 (\eta,\eta')$ could be
affected substantially by new physics masquerading as an electroweak penguin.  
Measurement of the $B_s \to \eta' \eta'$ and $B_s \to \eta \eta'$ branching
ratios could help resolve the question of whether the enhanced rate for $B \to
K \eta'$ decays is due in part to a singlet penguin contribution \cite{DGR} or
whether conventional penguin contributions suffice \cite{BN,HJLP}.
\bigskip

\centerline{\bf IX.  RELATED PUZZLES INVOLVING VECTOR MESONS}
\bigskip

\leftline{\bf A.  The parameter $S_{\phi K_S}$}
\bigskip

The time-dependent asymmetry parameter $S_{\phi K_S}$ in $B^0 \to
\phi K_S$ differs from the standard prediction of $\simeq 0.73$ by
about $1.8 \sigma$, as shown in Table \ref{tab:phiKasym}.  This
decay mode was one which was deemed promising for manifestation of
new physics in $b \to s$ penguin amplitudes well before any
measurements were made \cite{Grossman:1996ke}.

\begin{table}
\caption{Time-dependent CP asymmetry parameters for $B^0 \to \phi K_S$.
Errors on averages include scale factor $S = \sqrt{\chi^2}$.
\label{tab:phiKasym}}
\begin{center}
\begin{tabular}{c c c c c} \hline \hline
Parameter & BaBar~\cite{BaBarphiK} & Belle~\cite{BelleKpi} & $S$ & Average
\\ \hline
$S_{\phi K}$ & $0.50\pm0.25^{+0.07}_{-0.04}$ & $0.08\pm0.33\pm0.09$
 & 1.03 & $0.35 \pm 0.21$ \\
$A_{\phi K}$ & $0.00\pm0.23\pm0.05$ & $0.08\pm0.22\pm0.09$ & $<1$ &
 $0.04 \pm 0.17$ \\ \hline \hline
\end{tabular}
\end{center}
\end{table}

The penguin amplitude contributing to $S_{\phi K}$ is exclusively
a $b \to s s \bar s$ term.  Both color-suppressed and color-favored
matrix elements of this operator can contribute.  Thus, this process
becomes particularly worth while for identifying a specific four-quark
operator in which new physics is appearing.  Nonetheless, since
the discrepancy with the standard picture is less than $2 \sigma$,
speculation again seems premature.

A model-independent approach to studying an anomaly in $B^0\to\phi K_S$ 
was presented in~\cite{Chiang:2003jn}, using flavor SU(3) to normalize the 
amplitude of this process by the penguin amplitude dominating $B^+\to K^{*0}
\pi^+$. Explicit models of the space-time structure of new four-quark operators
for $b \to s q \bar q$~\cite{NPmodels} will in general treat $B \to VP$ decays 
(such as $B^0 \to \phi K_S$) differently from the $B \to PP$ 
decays which have occupied the bulk of our discussion.  This should be 
borne in mind when discussing possible deviations from the standard 
model in processes dominated by $b \to s$ penguin amplitudes.  This 
is in addition to any differences associated with the flavor $q$ in $b \to
s q \bar q$ amplitudes.  Thus, it is dangerous to quote average
values of $S_f$ when discussing different final states $f$.
\bigskip

\leftline{\bf B.  Helicity structure in $B^0 \to \phi K^{*0}$}
\bigskip

The $b \to s$ penguin amplitude (again, with Lorentz structure possibly
different from that in $B \to PP$ or $B \to VP$ decays) is expected to
dominate the process $B^0 \to \phi K^{*0}$.  In contrast, several other
processes with large branching ratios such as $B^0 \to \rho^+ \rho^-$ and
$B^\pm \to \rho^\pm \rho^0$ are expected to be dominated by the tree amplitude.
In these, the vector mesons appear to be almost totally longitudinally
polarized~\cite{Aubert:2003xc}, while the longitudinal fraction in $B^0 \to
\phi K^{*0}$ appears to be more like 1/2~\cite{Aubert:2004xc}.  Some authors
(see., e.g., Ref.\ \cite{Kagan:2004uw}) have cited this circumstance
as further evidence for the anomalous behavior of the penguin amplitude.

We see no reason why the penguin amplitude should have the same
space-time structure as the tree amplitude.  If, for example, it is
an effective operator driven partly by rescattering from charm-anticharm
states, as suggested in Ref.\ \cite{Colangelo:1989gi,SCET}, 
its space-time properties may be governed largely by long-distance
physics, and not amenable to the usual arguments based on Fierz 
rearrangement of a $V-A$ current.
\bigskip

\centerline{\bf X.  FUTURE EXPERIMENTAL TESTS}
\bigskip

It has sometimes been noted (see, e.g., Refs.\ \cite{Group:2004cx,%
Ligeti:2004ak,Giorgi}) that processes dominated by $b \to s$ penguin amplitudes
give an effective average $S_f$ value of about $0.4 \pm 0.1$, to be
contrasted with the Standard Model expectation of $\sin 2 \beta
\simeq 0.73$.  We regard this viewpoint as dangerous for three
reasons.  (1) It does not take account of the discrepancies between
the BaBar and Belle determinations of $S_f$ for several cases,
including that of $\eta' K_S$ mentioned above as well as $K_S
f_0(980)$, where $f_0(980) \to \pi \pi$~\cite{BelleKpi,f0}, and 
$K_SK_SK_S$~\cite{KsKsKs}.  When these discrepancies
are taken into account and the errors on experimental averages are
multiplied by an appropriate scale factor, the significance of the
deviation becomes less.  (2)  The several processes dominated by 
a $b\to s$ penguin amplitude involve small but different terms 
proportional to $V^*_{ub}V_{us}$~\cite{Gronau:2003kx,Gronau:2004hp,
Chiang:2003rb,Grossman:2003qp,Gronau:2003ep}. This implies 
{\it ab initio} different nonzero values for $\Delta S_f$ and $A_f$
for different final states $f$.  (3) As we have pointed out, the penguin
operators for $b \to s u \bar u$, $b \to s d \bar d$, and $b \to s s
\bar s$ may differ from one another, both in their instrinsic
strengths and in their matrix elements between states of different
spins.  How would one sort out this situation?

Our first suggestion is to concentrate on processes for which the
interpretation is as clean as possible.  Thus, $B^0 \to K^0 \pi^0$
appears considerably simpler to interpret in terms of specific
contributions than $B^0 \to \eta' K_S$, for which even the
interpretation of the decay rate itself has been the subject of
controversy.  (See, e.g., a discussion in Ref.\ \cite{Chiang:2004nm}).
Pinning down the value of $S_{K \pi}$ is a first priority. 
Lowering the experimental upper limit on ${\cal B}(B^0\to K^+K^-)$
may soon imply $S_{K \pi}\ne \sin 2\beta$, consistent with our prediction
(\ref{A_Kpi}) of a nonzero (negative) direct asymmetry $A_{K \pi}$.
Testing the sum rule (\ref{SR}), or determining whether $R_n$ differs 
from $R_c$ by a significant amount, obviously has high priority.

Our second suggestion regards the measurement of decay modes listed
in Table \ref{tab:ant}.  Improvement on the upper bounds listed
there to the level of bounds anticipated from SU(3) fits will not
only help sharpen bounds on $S_{\eta' K}$, but may uncover
additional unanticipated contributions to amplitudes or shortcomings
of the flavor SU(3) fits.

A third suggestion regards confirmation of the patterns of
tree-penguin interference seen in non-strange $B$ decays using
$B_s$ decays.  There are several $B_s$ decays related via U-spin to
$B^0$ decays, as noted in Section VIII.  Study of $B_s$ decays will
also be helpful in identifying the source of the enhanced rate for
$B \to \eta' K$.

A fourth suggestion is to continue the study of $B \to VV$ modes
which has begun so auspiciously with the study of such decays as
$B \to \rho \rho$, $B \to \phi K^*$, $B \to \rho K^*$, and
even $B_s \to \phi \phi$. Information on
these modes is approaching the stage that will permit analyses based
on flavor SU(3) analogous to those performed for $B \to PP$
\cite{Chiang:2004nm} and $B \to VP$ \cite{Chiang:2003pm} decays.
Relations among amplitudes have to be analyzed separately for each
helicity state, so it does not suffice to have rate information
alone.
\bigskip

\centerline{\bf XI.  SUMMARY AND CONCLUSIONS}
\bigskip

We have discussed several $B$ meson decay processes governed by the
$b \to s$ penguin amplitude, concentrating on processes with two
light pesudoscalar mesons $P$ in the final state.  Although several
indications appear for anomalous behavior, including the rate and
time-dependent asymmetry parameter $S$ for $B^0 \to K^0\pi^0$ and the
corresponding parameter $S$ for $B^0 \to \eta' K^0$, no deviations
of more than $2 \sigma$ from the standard model expectations have
been identified yet.  We have indicated several ways in which
experimental searches for this anomalous behavior can be sharpened.
\bigskip

\centerline{\bf ACKNOWLEDGMENTS}
\bigskip

We thank F. Blanc, H. Lipkin, D. Pirjol, A. Roodman, J. Smith and D. Suprun for
helpful discussions. J. L. R. wishes to acknowledge the generous hospitality of
the Technion during part of this study.  This work was supported in part by the
United States Department of Energy under Grant No.\ DE FG02 90ER40560,
by the Israel Science Foundation founded by the Israel Academy of Science
and Humanities, Grant No. 1052/04, and by the German--Israeli Foundation
for Scientific Research and Development, Grant No. I-781-55.14/2003.

\end{document}